\newcommand\etal{et~al.}
\newcommand\kms{\ifmmode {\rm\,km\,s^{-1}}\else${\rm\,km\,s^{-1}}$\fi}
\newcommand\simlt{\mathrel{\spose{\lower 3pt\hbox{$\mathchar"218$}} \raise 2.0pt\hbox{$\mathchar"13C$}}}
\newcommand\simgt{\mathrel{\spose{\lower 3pt\hbox{$\mathchar"218$}} \raise 2.0pt\hbox{$\mathchar"13E$}}}
\newcommand\aap{{\em A\&A}}

\newcommand\aj{{\em AJ}}
\newcommand\apj{{\em ApJ}}
\newcommand\apjl{{\em ApJ}}

\newcommand\mnras{{\em MNRAS}}
\newcommand\nat{{\em Nature}}

\documentclass[usenatbib]{mn2e}

\usepackage{psfig}
\usepackage{rotating}
\usepackage{amssymb}

\title[Spatial clustering of galaxies and distant radio galaxies]{Spatial clustering of USS sources and galaxies\thanks{Based on observations obtained with
the Australia Telescope Compact Array, the Anglo-Australian Telescope (Program
70.A-0514)}.}

\author[C. Bornancini \etal]{
\parbox[t]{\textwidth}{
Carlos~G.~Bornancini$^1$, Nelson~D.~Padilla$^2$, Diego~G.~Lambas$^{1,3}$
Carlos De Breuck$^4$}
\vspace*{6pt} \\ 
$^1$Grupo de Investigaciones en Astronom\'\i a Te\'orica y Experimental, IATE\\
Observatorio Astron\'omico, Universidad Nacional de C\'ordoba\\
Laprida 854, X5000BGR, C\'ordoba, Argentina.\\
$^2$Department of Astronomy, Pontificia Universidad Cat\'olica\\
 Vicu\~na Mackenna 4860, Casilla 306 Santiago 22, Chile\\
$^3$Consejo Nacional de Investigaciones Cient\'\i ficas y T\'ecnicas (CONICET),
Avenida Rivadavia 1917, C1033AAJ, Buenos Aires, Argentina.\\
$^4$ European Southern Observatory, Karl Schwarzschild Stra\ss e 2, D-85748 Garching, Germany.\\}
\pubyear{2005}
\begin{document}
\maketitle

\begin{abstract}
We present measurements of the clustering properties of galaxies in the field of redshift range $0.5 \lesssim z \lesssim 1.5$ Ultra Steep Spectrum (USS) radio sources selected from Sydney University Molonglo Sky Survey and NRAO VLA Sky Survey.
Galaxies in these USS fields were identified in deep near-IR observations, complete down to $K_s=20$, using {\tt IRIS2} instrument at the AAT telescope.
We used the redshift distribution of  $K_{s} < 20$ galaxies taken from Cimatti et al. (2002) to constrain the correlation length $r_0$.
We find a strong correlation signal of galaxies with $K_{s} < 20$ around our USS sample.
A comoving correlation length $r_{0}=14.0\pm2.8$ $h^{-1}$ Mpc and $\gamma=1.98\pm0.15$ are derived in a flat cosmological model Universe. 

We compare our findings with those obtained in a cosmological N--body simulation populated with GALFORM semi-analytic galaxies. 
We find that clusters of galaxies with masses in the range $M=10^{13.4-14.2}$ $h^{-1}$ M$_{\sun}$ have a cluster--galaxy cross--correlation
amplitude comparable to those found between USS hosts and galaxies.
These results suggest that distant radio galaxies are excellent tracers of galaxy overdensities and pinpoint the progenitors of present day rich clusters of galaxies.

\end{abstract}


\section{Introduction}

In hierarchical galaxy formation models, cosmic structures form by the gravitational amplification of small primordial fluctuations of matter density in the early Universe \citep{white}.
Studies of the clustering properties of galaxies at high redshift are essential for understanding galaxy and structure formation.
One of the most commonly used statistics to measure the clustering of
a population of sources is the two-point correlation function
$\xi(r)$, which measures the excess probability of finding a pair of
objects at a separation $r$ with respect to a random distribution.

High redshift radio galaxies are ideal targets for pinpointing massive systems. 
Radio galaxies follow a close relation in the Hubble $K-z$ diagram \citep{lilly84}.
The nature of this behavior in the $K-z$ diagram shows that the stellar luminosities of  $z\gtrsim 1$ radio galaxies are more luminous than normal galaxies at these redshifts  \citep{deb02}. 
At lower redshifts they are found frequently in moderately rich clusters \citep{hill,yates}
Recently, galaxy overdensities comparable to that expected for clusters of Abell class 0 richness are found near radio galaxies up to $z=1.6$ \citep{best2000,best2003,bornan}.There is increasing evidence that galaxy overdensities around radio galaxies existed at very high redshifts. Using Ly$\alpha$ and/or H$\alpha$ techniques, \citet{kurk2000,vene2002,mileynat} found overdensities of companion galaxies around powerful $2<z<4$ radio galaxies.

In this paper we estimated the spatial correlation length for galaxies in the fields of USS targets selected from the 843 MHz Sydney University Molonglo sky Survey (SUMSS) and 1.4 GHz NRAO VLA Sky Survey (NVSS) in the redshift range $0.5 \lesssim z  \lesssim 1.5$, through the Limber's equation using an appropriate observed redshift distribution. 
We compared our results with those obtained in cosmological N--body simulations.   
This paper is organized as follows: Section 2 describes the sample analyzed, we investigate the USS-galaxy cross-correlation analysis in Section 3. In Section 4 we interpret our results with those obtained in a cosmological N--body simulation.
Finally we discuss our results in Section 5.

Throughout this paper we will use a flat cosmology with density parameters $\Omega_{M}=0.25$, $\Omega_{\Lambda}=0.75$ and a Hubble constant $H_{0}=100$ $h$ km s$^{-1}$ Mpc$^{-1}$.

\section{The Data}
The USS sample selection, radio data and redshifts used for this analysis was presented and described by 
\citet{sumss,sumss2}.
Detailed descriptions of the construction of the galaxy catalogue is given in \citet{yo}.
In summary, we used 11 $K_s$--band images centered in Ultra Steep Spectrum (USS) radio sources selected from the Sydney University Molonglo sky Survey and NRAO VLA Sky Survey, obtained with the instrument {\tt IRIS2} at the AAT telescope.
We selected the redshift of the USS sample in the range $0.5 \lesssim z \lesssim 1.5$.
The redshift upper limit was adopted in order to assure that $K$--band images are sufficiently deep while the lower limit precludes the use of too close USS for which the field of view subtends too small linear scales.    

Our sample sources is listed in Table~1 in IAU J2000 format, including the $K_{s}$ counterpart magnitude and the spectroscopic redshift \citep{sumss2}.

\section{USS-GALAXY CROSS-CORRELATION ANALYSIS}

The spatial USS-galaxy cross-correlation function $\xi _{ug}(r)$ is
defined as the excess  probability $dP$ of finding a galaxy in the volume
element $dV$ at a distance $r$ from a USS target,
 
\begin{equation}
dP=\bar{n}\left[ 1+\xi _{ug}(r)\right] dV,  
\label{dP}
\end{equation}%
where $\bar{n}$ is the mean space density of galaxies. 
The spatial two-point cross-correlation function $\xi _{ug}(r)$ has been shown to be well approximated by a power law:

\begin{equation}
\xi _{ug}(r)=\left(\frac{r}{r_{0}}\right)^{-\gamma}.
\label{xi}
\end{equation}%

In order to obtain the cross--correlation length $r_{0}$ we first determine the projected cross-correlation function $\omega _{ug}(\sigma)$, where $\sigma$ is the projected separation between a USS target and a galaxy at redshift $z$.

We use the following estimator of the projected cross-correlation function, \citep{peebles80}:

\begin{equation}
\omega(\sigma)=\frac{n_R}{n_G}\frac{UG(\sigma)}{DR(\sigma)}-1,
\end{equation}
where $n_G$ and $n_R$ are the numbers of galaxies in the sample and in a random sample respectively, $UG(\sigma)$ is the number of real USS-galaxy pairs separated by a projected distance in the range $\sigma$, $\sigma+\delta \sigma$,
and $DR(\sigma)$ are the corresponding pairs when considering the random
galaxy sample. 
We estimate the corresponding correlation length using the Limber equation \citep{limber}.
The power law model for $\xi _{ug}(r)$ gives:

\begin{equation}
\omega(\sigma)=B \sqrt{\pi} \frac{\Gamma \left[(\gamma-1)/2\right] }{\Gamma(\gamma/2)}\frac{r_{0}^{\gamma}}{\sigma^{\gamma-1}},
\label{lil}
\end{equation}
where the constant $B$ regulates the amplitude of the correlation function taking
into account the differences in the selection function of USS targets and
galaxies and can be calculated by \citep{lilje}:
 
\begin{equation}
B=\frac{\sum_{i} N(y_i)} {\sum_{i}\frac{1}{y_{i}^2}\int_{0}^{\infty}%
N(x)x^2dx},  
\label{b}
\end{equation}
\setcounter{figure}{0}
\begin{figure}
\begin{tabular}{ll}
\psfig{file=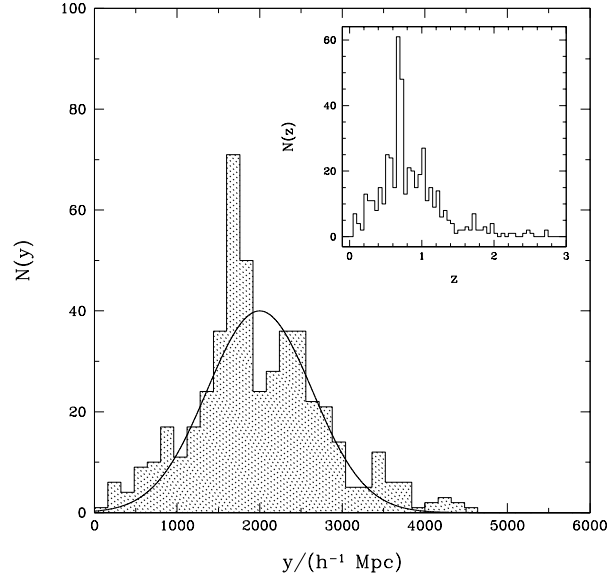,width=8cm}
\end{tabular}
\caption{Distance distribution for $K_{s} < 20$ galaxies taken the K20 survey \citep{cima}.
The dashed curve shows the selection function obtained fitting a Gaussian function. The upper small box shows the corresponding redshift distribution of the sample.}
\label{distri}
\end{figure}
where $N(y_i)$ is the selection function of the galaxy survey, $y_i$ is the
distance to USS target $i$ and the sum extends over all USS targets in the sample.
Equation \ref{lil} can be easily solved analytically if we perform a
linear interpolation of $\omega _{ug}$ between its values at the measured $\sigma $'s.
In order to calculate the constant $B$ we evaluated the selection function $N(y_i)$ using the redshift (spectroscopic and photometric) distribution of $K_{s} < 20$ galaxies published by \citet{cima}\footnote{Data and further
information available at \texttt{http://www.arcetri.astro.it/$\sim$k20/releases}}.
We model the distance distribution using a Gaussian function, and find that $\bar y=2000$  $h^{-1}$ Mpc (mean) and $\sigma_y=635$  $h^{-1}$ Mpc (standard deviation) is a reasonable set of values that reproduces the distribution (See Figure~\ref{distri}).
In Figure \ref{pro} we show the projected cross-correlation function $\omega_{ug}(\sigma)$ for USS targets with spectroscopic redshifts in the range $0.6 \lesssim z \lesssim 1.5$ and galaxies with $K_{s}<20$.
We estimate cross-correlation function error bars using the $jackknife$ technique \citep{efron}.
We find a comoving correlation length $r_0=14.0\pm2.8 $ $h^{-1}$ Mpc with slope $\gamma=1.98\pm$0.04.

We tested the accuracy of this result by varying the $N(y_i)$ distribution over a reasonable 
range of values for $\bar y$ and $\sigma_y$. We find that the calculated $r_{0}$ is only weakly 
dependent on the $\bar y$ used, and is only affected at the 10\% level. 

\setcounter{figure}{1}
\begin{figure}
\begin{tabular}{ll}
\psfig{file=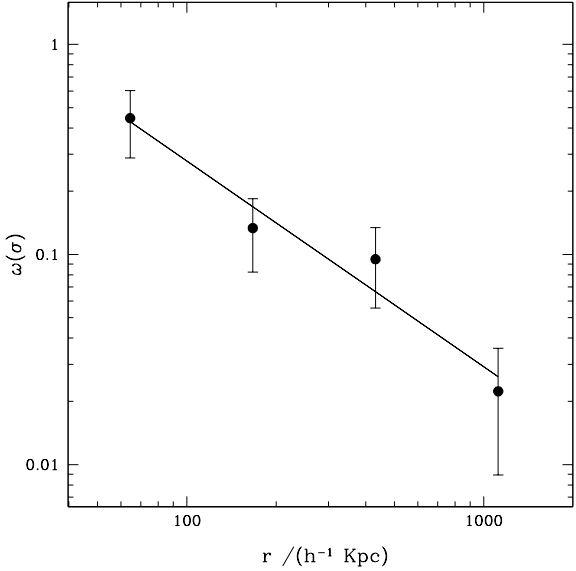,width=8cm}
\end{tabular}
\caption{Projected cross-correlation function $\omega_{ug}(\sigma)$ for USS targets with spectroscopic redshift in the range $0.6 \lesssim z \lesssim 1.5$. The error bars correspond to the 1 $\sigma$ uncertainty estimated using the jackknife technique.}
\label{pro}
\end{figure}

\section{Comparison with N--body simulations}
We interpret our results with the aid of a cosmological N--body simulation populated
with GALFORM semi-analytic galaxies \citep{cole} at different outputs corresponding to different redshifts,
$z=0,1$, and $3$.  This simulation was kindly provided by the Durham group. 
The cosmological model corresponds to matter and cosmological constant density 
parameters $\Omega_m=0.25$, $\Omega_{\Lambda}=0.75$, a power spectrum tilt $n=0.95$,
an amplitude of fluctuations of $\sigma_8=0.8$, and a Hubble constant of $H_{0}=100$ $h$ kms$^{-1}$ Mpc$^{-1}$, where
$h=0.7$.  The total number of particles is $1080^3$, the mass resolution is $5.05\times10^{10}$h$^{-1}$M$_{\sun}$,
and the number of dark matter haloes with masses greater than $M=10^{12}$h$^{-1}$M$_{\sun}$
ranges from $\sim400,000$ at $z=3$ to $\sim2,200,000$ at $z=0$.
The number of GALFORM galaxies ranges from $\sim20,000,000$ to $120,000,000$ at $z=3$ and $z=0$ respectively.

We now briefly explain the procedure by which galaxies are assigned their 
properties
in the Semi-analytic code.  GALFORM is run for each halo in the numerical
simulation, where galaxies
are assigned to a randomly selected dark matter halo particle.  The
different galaxy properties
such as magnitudes in different bands, including the $K$--band, depend on the
dark matter halo
merger tree.  This merger tree is generated via Monte-Carlo modelling
based on the
extended Press-Schechter theory, and the evolution of the galaxy
population in the halo
is followed through time and different processes are considered in this
evolution, including
gas cooling, quiescent star formation and star formation bursts, mergers,
galactic winds
and super winds, metal enrichment, extinction by dust.  For full details
on the modelling
the reader is referred to \citet{cole}.

We calculate the cross--correlation function using the simulation haloes with masses above a lower 
mass limit as centres, and as tracers, the GALFORM semi-analytic galaxies.  By comparing
these measurements to the results from the cross--correlation between USS and normal galaxies, we make
the implicit assumption that USS galaxies reside at the centres of dark-matter haloes.  This comparison
will make it possible to infer the mass of the structures associated to the USS hosts.  

Figure \ref{nel} shows the resulting
real-space cross-correlation functions between haloes and semi-analytic galaxies at $z=1$ (top panel) for
different halo masses.  The shaded area corresponds to the power law fit for the real-space
correlation function inferred from the cross--correlation function
for USS radio sources with spectroscopic redshifts in the range $0.6 \lesssim z \lesssim 1.5$ (See Figure 2). 
In the middle panel of this figure, we compare the values of USS--galaxy
cross-correlation length as a function of halo mass for three different redshift outputs from the numerical
simulations; as can be seen, the observed values are consistent with cluster masses 
within $M=10^{13.4-14.2}$ $h^{-1}$ M$_{\sun}$ at redshift z=1, 
indicating that our USS sample resides in massive clusters.  
In order to check whether our observational estimate of $r_0$ is affected by systematic biases,
we calculate the projected correlation function in the numerical simulation and recover the
real-space correlation length using Eq. \ref{lil}, setting $B=1$; we consider the same range of
separations available in the real data.  The results for $z=1$ are shown in filled circles in
the middle panel.  As can be seen, our conclusions on the mass of USS host haloes changes only
slightly to $M=10^{13.2-13.8}$ $h^{-1}$ M$_{\sun}$,
although we note that for lower and higher halo masses, the value of $r_0$ recovered from projected correlations
is underestimated and overestimated, respectively. 
This test provides a useful check of our observational results 
which were derived from relatively small projected scales 
($r_p \lesssim$1 Mpc). Our findings in the simulations indicate 
that reliable $r_0$ values are obtained using the power law approximation 
applied to projected correlations for $r_p <1 h^{-1}$ Mpc when the true correlation length is lesser than $\sim 15 h^{-1}$ Mpc corresponding to host halo mass $M\lesssim$ 10$^{14}$ M$_{\sun}$.

A further indication of the
mass of USS galaxy host haloes comes from the lower panel of this figure, where the lines
correspond to the projected cross-correlation function measured in the GALFORM simulation for
different masses (High to low masses from top to bottom lines at $\log_{10}(\sigma/$h$^{-1}$Mpc$)=-0.3$).  
$\Xi(\sigma)$ is calculated directly using,

\begin{equation}
\Xi(\sigma)={\rm Norm.}\ 2\int_{0}^{\pi_{max}}\xi(\sqrt{\sigma^2+\pi^2}){\rm d}\pi,
\end{equation}

where we have used $\pi_{max}=80$ h$^{-1}$Mpc, and the normalisation, ${\rm Norm.}$, is set so that $\Xi(\sigma)$ and
$\omega(\sigma)$ coincide at $\log_{10}(\sigma/$h$^{-1}$Mpc$)=-1$.  The gray area shows the measured values
of $\omega(\sigma)$ from the USS sample; as can be seen the measured projected correlation function is in best agreement for $M\sim10^{13.85}$h$^{-1}$M$_{\sun}$.

\setcounter{figure}{2}
\begin{figure}
\begin{tabular}{ll}
\psfig{file=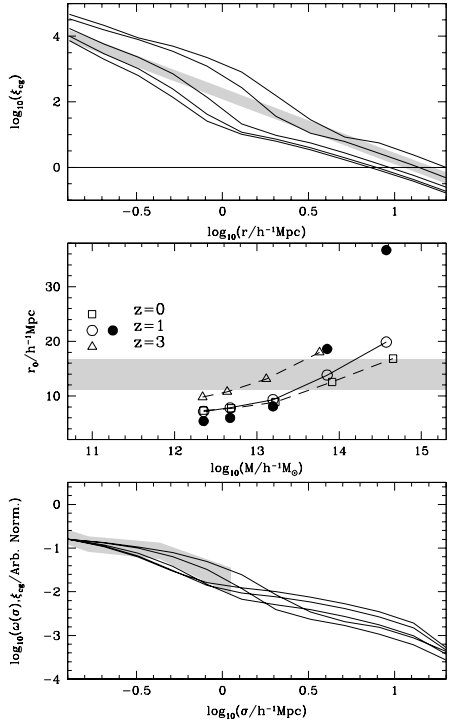,width=11cm}
\end{tabular}
\caption{
Top panel: cluster-galaxy cross-correlation functions from the simulation
output at $z=1$ for increasing halo masses $Log(M)=12.35, 12.68, 13.2, 13.85$~and~$14.57$ $h^{-1}$ $M_{\sun}$ (bottom to top).
The shaded area
shows the real--space correlation function measured from the USS targets with spectroscopic redshift in the range $0.6 \lesssim z \lesssim 1.5$.
Middle panel: the real-space correlation length, $r_0$, as a function
of halo mass for three different redshift outputs; the open symbols show the values recovered
from the real-space correlation function, whereas the filled symbols (shown only
for $z=1$) indicate the values of $r_0$ obtained from the projected correlation
function following a similar procedure to that applied to the real data.  The shaded area shows the
allowed range of $r_0$ for the USS results.
Bottom panel: the projected USS cross-correlation function (gray area) and 
$\Xi(\sigma)$ from the GALFORM simulation for the same halo masses as in the top panel; the normalisation
is set so that $\Xi(\sigma)$ and $\omega(\sigma)$ coincide at $\log_{10}(\sigma/$h$^{-1}$Mpc$)=-1$. 
}
\label{nel}
\end{figure}

\section{Conclusions }

We have analyzed the clustering properties of galaxies in the field of $0.5 \lesssim z \lesssim 1.5$ Ultra Steep Spectrum (USS) radio galaxies selected from the Sydney University Molonglo Sky Survey 
(SUMSS) and NRAO VLA Sky Survey (NVSS).
We estimated the spatial clustering correlation length for galaxies in these fields, using the Limber equation using an appropriate observed redshift distribution, and we examined the dependence of galaxy clustering on USS targets luminosity.
A comoving correlation length $r_0=14.0\pm2.8$ $h^{-1}$ Mpc is derived and a slope $\gamma=1.98\pm$0.15.

From our comparison with numerical simulations, we find that clusters of galaxies with masses in the range $M=10^{13.4-14.2}$ $h^{-1}$ M$_{\sun}$ have a cluster--galaxy correlation amplitude comparable to that found between USS hosts and galaxies.
Our testing with the numerical simulations also indicate that these observational results are not severely affected by the relatively small projected scales 
explored. We notice that for larger spatial correlation lengths, 
the power-law extrapolation for observed projected correlations 
in $\sim 1 h^{-1}$Mpc fields would not give confident results. 

Previous studies have addressed the clustering of galaxies around radio sources.
\citet{wold}, using different cosmological parameters
obtained a radio quasar-galaxy cross-correlation length
about twice as large as the local galaxy autocorrelation length 
in suitable agreement with our findings. The more recent work by \citet{barr} also indicates a moderate galaxy density enhancement around radio loud
quasars similar to Abell richness 0 clusters. 

Our analysis suggest that distant luminous radio galaxies are excellent tracers of galaxy overdensities and may pinpoint the progenitors 
of present day rich and moderate rich clusters of galaxies.

\section{Acknowledgements}
The authors are specially grateful to C. Baugh and the Durham group for providing the GALFORM
semi--analytic simulation output used in this work.
C. Bornancini thanks to Julian Mart\'{\i}nez for helpful comments and suggestions.
This work was supported in part by the ESO-Chile Joint Committee, NDP 
was supported by a Proyecto Postdoctoral Fondecyt no. 3040038.  
This work was partially supported by the
Consejo Nacional de Investigaciones Cient\'{\i}ficas y T\'ecnicas,
Agencia de Promoci\'on de Ciencia y Tecnolog\'{\i}a,  Fundaci\'on Antorchas
and Secretaria de Ciencia y
T\'ecnica de la Universidad Nacional de C\'ordoba, and the European Union Alfa II Programme, through LENAC, the Latin American--European Network for Astrophysics and Cosmology

\setcounter{table}{1}
\begin{onecolumn}
\label{t0}
\small
Table 1. USS Sample characteristics. Designation in IAU J2000 format, together with the $K_{s}$ counterpart magnitude, and the spectroscopic redshift (De Breuck et al, in prep.).
\begin{center}
\begin{tabular}{lccccc}
\hline
(1) &(2)&(3)\\
Name & K$-$mag.& $z$ \\
& $_{\tt MAG\_BEST}$    &spectroscopic \\
\hline
NVSS~J015232$-$333952  & 16.26$\pm$0.02 &0.6148$\pm$0.001    \\ 
NVSS~J015544$-$330633  & 16.93$\pm$0.05 &1.048$\pm$0.002     \\ 
NVSS~J021716$-$325121  & 18.75$\pm$0.20 &1.384$\pm$0.002     \\ 
NVSS~J030639$-$330432  & 17.88$\pm$0.11 &1.201$\pm$0.001     \\ 
NVSS~J202026$-$372823  & 18.56$\pm$0.15 &1.431$\pm$0.001     \\ 
NVSS~J204147$-$331731  & 16.86$\pm$0.05 &0.871$\pm$0.001     \\ 
NVSS~J225719$-$343954  & 16.53$\pm$0.02 &0.726$\pm$0.001     \\ 
NVSS~J230203$-$340932  & 17.34$\pm$0.07 &1.159$\pm$0.001     \\ 
NVSS~J231519$-$342710  & 18.10$\pm$0.13 &0.970$\pm$0.001     \\ 
NVSS~J234145$-$350624  & 15.94$\pm$0.04 &0.641$\pm$0.001     \\ 
NVSS~J234904$-$362451  & 17.63$\pm$0.14 &1.520$\pm$0.003     \\ 
\hline
\end{tabular}
\end{center}
\end{onecolumn}

{}

\end{document}